\documentclass[12pt]{iopart}

\usepackage{iopams}
\usepackage{amssymb}
\usepackage{bm}
\usepackage{amsfonts}
\usepackage{graphicx,epsfig}
\usepackage{graphicx,color}

\begin{document}

\title{$N$-photon bundles emission in high-spin Jaynes-Cummings model}
\author{Huanhuan Wei}
\address{Guangdong Provincial Key Laboratory of Quantum Metrology and Sensing $\&$ School of Physics and Astronomy, Sun Yat-Sen University (Zhuhai Campus), Zhuhai 519082, China}

\author{Jing Tang}
\address{School of Physics and Optoelectronic Engineering, Guangdong University of Technology, Guangzhou 510006, China}
\address{Guangdong Provincial Key Laboratory of Sensing Physics and System Integration Applications, Guangdong University of Technology, Guangzhou, 510006, China}

\author{Yuangang Deng}
\address{Guangdong Provincial Key Laboratory of Quantum Metrology and Sensing $\&$ School of Physics and Astronomy, Sun Yat-Sen University (Zhuhai Campus), Zhuhai 519082, China}
\ead{jingtang@gdut.edu.cn}
\ead{dengyg3@mail.sysu.edu.cn}
\vspace{10pt}


\begin{abstract}
High-spin quantum systems, endowed with rich internal degrees of freedom, constitute a promising platform for manipulating high-quality $n$-photon states. In this study, we explore $n$-photon bundles emission by constructing a high-spin Jaynes-Cummings model (JCM) within a single-mode cavity interacting with a single spin-$3/2$ atom. Our analysis reveals that the $n$-photon dressed state splittings can be significantly enhanced by adjusting the linear Zeeman shift inherent to the internal degrees of freedom in high-spin systems, thereby yielding well-resolved $n$-photon resonance. The markedly enhanced energy-spectrum anharmonicity, stemming from strong nonlinearities, enables the realization of high-quality $n$-photon bundles emission with large steady-state photon numbers, in contrast to conventional spin-1/2 JCM setups. Of particular interest is the realization of an optical multimode transducer capable of transitioning among single-photon blockade, two- to four-photon bundles emission, and photon-induced tunneling by tuning the light-cavity detuning in the presence of both cavity and atomic pump fields. This work unveils significant opportunities for diverse applications in nonclassical all-optical switching and high-quality multiphoton sources, deepening our understanding of creating specialized nonclassical states and fundamental physics in high-spin atom-cavity systems.
\end{abstract}


\section{Introduction}
Multiphoton states, serving as fundamental elements of quantum lights, provide versatile applications in quantum communication~\cite{Nature.35106500}, quantum computing~\cite{Nature35005001,RevModPhys.79.135}, quantum metrology~\cite{PhysRevLett.96.010401, RevModPhys.90.035005} and fundamental examinations of quantum physics~\cite{RevModPhys.90.035006, Lei2023}. A pivotal means to  generate these states is through $n$-photon blockade (PB)~\cite{PhysRevLett.122.123604}, which is known as nonlinear quantum scissors~\cite{Prog.Opt.56.131}. This effect transforms classical light into non-classical light, by impeding the generation of the subsequent photons when $n$ photons are present, drawing substantial attention due to its potential applications in quantum information and communication.

Creating single-PB relies on two typical physical mechanisms. One is the conventional PB, with requiring a sufficiently large energy-spectrum anharmonicity~\cite{PhysRevA.49.R20, PhysRevLett.109.193602}. The other is unconventional PB, operating via destructive quantum interference among various quantum transition pathways~\cite{PhysRevLett.104.183601, PhysRevA.83.021802, PhysRevLett.108.183601}. Both these PB types have been extensively investigated across various quantum systems, encompassing cavity quantum electrodynamics (QEDs) systems~\cite{Nature.03804, Nature.07112, Tang:21, tang2015, PhysRevA.93.013856}, superconducting circuit QED systems~\cite{PhysRevLett.107.053602, PhysRevA.89.043818}, Kerr-type nonlinear cavities~\cite{PhysRevA.82.013841, PhysRevLett.123.013602, PhysRevA.91.063808, PhysRevA.96.053810}, and optomechanical systems~\cite{PhysRevLett.107.063601, PhysRevLett.107.063602, PhysRevA.99.043837, Li:19}. At the same time, $2$-PB has been theoretically proposed and experimentally advanced in Kerr-type system~\cite{PhysRevA.87.023809,PhysRevLett.118.133604}, strong-coupling qubit-cavity system~\cite{PhysRevA.91.043831, SHAMAILOV2010766}, and a cascaded cavity QED system~\cite{PhysRevLett.118.133604, PhysRevA.98.043858}. Furthermore, investigations into $n$-PB $(n>2)$ have also been explored~\cite{PhysRevA.95.063842,PhysRevA.99.053850,PhysRevA.90.013839, PhysRevLett.121.153601}. The mechanism for generating $n$-PB relies on the great strong energy-spectrum anharmonicity. Notably, achieving $n$-PB through the unconventional mechanism, involving intricate conditions of destructive quantum interference, poses significant challenges.

In contrast to the extensively studied $n$-PB, the emission of $n$-photon bundles, representing a new building block in quantum optics, releases bundles of strongly correlated photons, offering opportunities for exploring fundamental physics and applications in quantum information science and technology~\cite{science.1188172, PhysRevLett.87.013602, nature07127, nphoton.2012.336}. The $n$-photon bundles states exhibit strong bunching for single photons and antibunching for separate bundles of photons. Mechanisms for generating $n$-photon (phonon) bundles include $n$-phonon resonance processes~\cite{Deng:21}, Mollow physics~\cite{PhysRevA.44.7717, Munoz2014, SanchezMunoz18, lpor.201700090}, deterministic parametric down-conversion~\cite{PhysRevLett.117.203602, PhysRevResearch.4.013013, qute.202000132}, Stokes multiphonon processes~\cite{PhysRevLett.124.053601} and parity-symmetry-protected multiphoton processes~\cite{PhysRevLett.127.073602}. These significant advances have predominantly focused on generating high-purity $n$-quanta states utilizing the standard two-level Jaynes-Cummings model (JCM) under strong coupling conditions, posing experimental challenges in cavity QEDs. Recently, high-quality non-classical single-photon to two-photon bundles emission has been proposed in a spin-1 JCM~\cite{tangspin1}. This proposal benefits from the large internal degrees of freedom, overcoming experimental constraints associated with strong atom-cavity coupling. Leveraging the internal degrees of freedom in high-spin systems, this mechanism opens new avenues for creating exotic nonclassical states and offers versatile applications in quantum information and metrology.

In this work, we propose to the generation of $n$-photon bundles emission utilizing a single atom trapped in a single-mode optical cavity with constructing a spin-$3/2$ JCM. We demonstrate the $n$-photon resonance can be finely tuned by manipulating the linear Zeeman shift, facilitated by the internal degrees of freedom inherent in the high-spin atom-cavity system. By independently driving the cavity field and atom pump field, high-quality single-PB and two-photon bundles emission are observed, respectively.  We show that the generalized second-order correlation functions for single-photon and two-photon bundles states can reach values as low as $g_{1}^{(2)}(0)\thickapprox7.3\times10^{-4}$ and $g_{2}^{(2)}(0)\thickapprox6.6\times10^{-4}$, respectively. These results indicate strong sub-Poissonian photon statistics for single photons and isolated two-photon bundles. Remarkably, we demonstrate the feasibility of an optical multimode transducer capable of transitioning among various photon states: ranging from single-PB to multiphoton bundles, as well as photon-induced tunneling (PIT). This broadens the horizon for diverse technological applications, spanning from nonclassical multimode transducers to high-quality multiphoton sources~\cite{s41377, s41586, science.aba}.


In contrast to the quadratic Zeeman shift utilized in spin-1 JCM~\cite{tangspin1}, the novelty lies in the distinct physical mechanism governing the generation of strong energy-spectrum anharmonicity leveraging the linear Zeeman shift. Thus we achieve three-photon and four-photon bundles emission with large steady-state photon numbers under simultaneous action of cavity-driven and atom-pump fields. These high-quality $n$-photon bundles state showcasing distinct statistical properties of photon bunching for single photons with $g_{1}^{(2)}(0)> g_{1}^{(2)}(\tau)$ and antibunching for separated ones with $g_{n}^{(2)}(0)< g_{n}^{(2)}(\tau)$, play a pivotal role in quantum metrology~\cite{PhysRevLett.96.010401,RevModPhys.90.035005}, quantum lithography and cryptography~\cite{PhysRevLett.87.013602, RevModPhys.74.145}. Compared to creating $n$-photon bundle states typically relying on Mollow physics~\cite{PhysRevA.44.7717, Munoz2014, SanchezMunoz18, lpor.201700090}, which depend on high-order processes operated in the far dispersive regime, our proposal benefits from the large internal degrees of freedom inherent to introduce a novel mechanism for significantly enhancing energy-spectrum anharmonicity in high-spin systems.

The paper is structured as follows: In Sec.~\ref{model and Hamiltonian}, we introduce the model and Hamiltonian responsible for realizing $n$-photon bundles emission in the spin-3/2 JCM system. In Sec.~\ref{energy spectrum}, we calculate the energy spectrum of the system. Sec.~\ref{simulation method} outlines the conditions necessary for achieving $n$-photon bundles emission. Sec.~\ref{numerical results} showcases the outcomes of numerical simulations concerning $n$-photon bundles emission. Finally, Sec.~\ref{conclusion} offers a concise summary of our results.

\begin{figure}[ht]
\centering
\includegraphics[width=0.99\columnwidth]{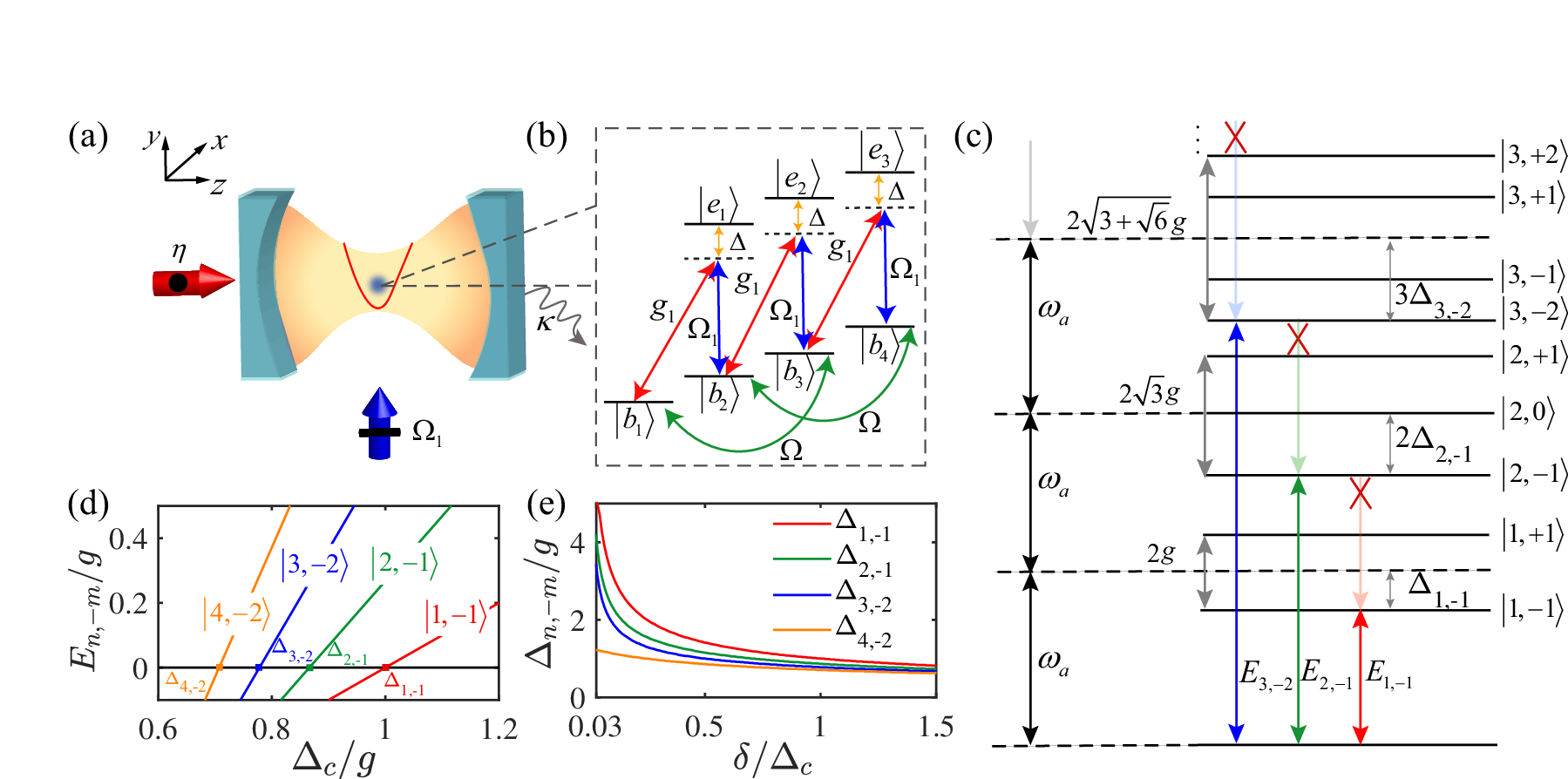}
\caption{(color online).  (a) Sketch of a single atom inside a high-finesse optical cavity for generating $n$-photon bundles emission. (b) Level diagram for a single spin-$3/2$ atom. (c) Anharmonicity energy spectrum of Hamiltonian~(\ref{eq:H2}) for $\delta=\Delta_c$. The energy splitting of the $n$th dressed states $|n, \pm m\rangle$ is determined using the analytic expression in Eq.~(\ref{splitting}). (d) $\Delta_c$ dependence of energy spectrum $E_{n,-m}$ with square denoting the positions of $n$-photon resonance $\Delta_{n, -m}$ $(m=1,2)$. (e) Variation of $n$-photon resonance $\Delta_{n, -m}$ $(m=1,2)$ as a function of $\delta$.} \label{fig:model}
\end{figure}%

\section{Model and Hamiltonian}
\label{model and Hamiltonian}
To generate $n$-photon bundles emission, we consider a high-finesse single alkaline-earth-metal (like) atom confined in a single-mode optical cavity, driven by a weak laser field with amplitude $\eta$ and frequency $\omega_L$, as depicted in Fig.~\ref{fig:model}(a). The atomic energy-level structure for the pseudo-spin $3/2$ atom is illustrated in Fig.~\ref{fig:model}(b). This structure our internal magnetic Zeeman levels of the ground states $|b_{j}\rangle$ with $j=1,2,3,4$ and three electronically excited states $|e_{j}\rangle$ with $j=1,2,3$. A bias magnetic field $\mathbf{B}$ along the cavity axis is defined as the quantization axis ($z$-axis), induces a significant Zeeman shift $\hbar \omega_{b}$ within the ground state manifold.

Regarding the optical cavity, its bare frequency and decay rate are denoted as $\omega_c$ and $\kappa$, respectively. In our scheme, a $\sigma^+$-polarized cavity field is far-off resonant with the atomic transitions $|b_{j}\rangle\leftrightarrow|e_{j}\rangle$, corresponding to the single-atom cavity coupling strength $g_{1}$ and atom-cavity detuning $\Delta$. To generate the Raman coupling, the atomic transitions $|b_{j+1}\rangle\leftrightarrow|e_{j}\rangle$ are coupled by a $\pi$-polarized classical pump field perpendicular to the cavity axis with Rabi frequency $\Omega_{1}$. The magnetic quantum numbers of the electronic states satisfy $m_{|b_{j}\rangle}=m_{|e_{j}\rangle}-1$ and $m_{|b_{j+1}\rangle}=m_{|e_{j}\rangle}$. Additionally, we consider the next nearest neighbors of ground states through a pair of $\sigma$-polarized classical laser fields, inducing the atomic transitions $|b_{j}\rangle\leftrightarrow|b_{j+2}\rangle$ with a weak Rabi frequency of $\Omega$, as illustrated in Fig.~\ref{fig:model}(b).  The frequency difference between the pair laser is adjusted to compensate for the Zeeman shift between $|b_{j}\rangle$ and $|b_{j+2}\rangle$ states. It is important to note that, in this laser configuration, the laser frequency responsible for the effective atomic pumping is significantly detuned from the   cavity resonance frequency.

In the case of large atom-cavity detuning limit, $g_1/\Delta\ll1$ and $\Omega_1/\Delta\ll1$, the three far-off resonance excited states $|e_{j}\rangle$ with $j=1,2,3$ can be adiabatically eliminated. Using the rotating-wave approximation, we can formulate the Hamiltonian describing the spin-3/2 JCM system along with the driving terms corresponding to the cavity field and the atomic pump field (refer to \ref{A}).

\begin{eqnarray} \label{eq:H1}
\hat{H}_1/\hbar&= \Delta_{c}\hat{a}^{\dagger}\hat{a}+\Delta_{a}\sum \limits _{j=1} ^{3} j \hat{b}_{j+1}^{\dagger}\hat{b}_{j+1}+ \eta(\hat{\emph a}^{\dagger}+\hat{\emph a}),\nonumber
\\& +\left[ g\hat{a}^{\dagger}\sum \limits _{j=1} ^{3}\hat{b}_{j}^{\dagger}\hat{b}_{j+1}+\Omega\sum \limits_{\emph j=1} ^{2}\hat{b}_{j}^{\dagger}\hat{b}_{j+2}+\rm H.c.\right],
\end{eqnarray}
where $\hat{a}^\dag$($\hat{a}$) is the creation (annihilation)
operator for the cavity mode and $\hat{b}_{j}^{\dagger}$ ($\hat{b}_{j}$) represents the atomic spin projection operator for the four relevant states with $j=1,2,3,4$. $\Delta_{c}=\Delta_{c}^{'}-{g_{1}^{2}}/{\Delta}$ is the effective cavity-light detuning,
$\Delta_{a}=\omega_{b}-|\Omega_{1}|^2/\Delta$ is the tunable effective single photon detuning of atom, and $g=-{g_{1}\Omega_{1}}/{\Delta}$ is the effective atom-cavity coupling.

To gain more insight of photon emissions, the spin-$3/2$ representation of the Hamiltonian~(\ref{eq:H1}) with neglecting the weak driving fields can be reformulated as:
\begin{eqnarray} \label{eq:H2}
\hat{H}_{2}/\hbar  =	\Delta_{c}\hat{a}^{\dagger}\hat{a}+\delta\hat{S}_{z}+g(\hat{a}\hat{F}_{+}+\hat{a}^{\dagger}\hat{F}_{-}).
\end{eqnarray}

Here $\delta=\Delta_{a}$ is the effective linear Zeeman shift and $\hat{S}_{z}$ denotes the spin-$3/2$ matrices with the angular momentum operator defined as
$\hat{S}_{z}=\sum \limits _{j=1} ^{4}(j-\frac{5}{2})\hat{b}_{j}^{\dagger}\hat{b}_{j}$.
For convenience of representation, the atomic transition matrix is denoted by $\hat{F}_{+}\left(\hat{F}_{-}\right)$, where the raising operator is expressed  as $\hat{F}_{+}=\sum \limits _{j=1} ^{3}\hat{b}_{j+1}^{\dagger}\hat{b}_{j}$ with $\hat{F}_{+}=\hat{F}_{-}^{\dagger}$.

\section{Energy spectrum of the system}
\label{energy spectrum}
To explore the physical processes of multiphoton bundles emission, the energy spectrum of the system is calculated by diagonalizing Hamiltonian in Eq.~(\ref{eq:H2}). The total number of excitations operator $\hat{N}=\hat{a}^{\dagger}\hat{a}+\hat{S}_{z}+3/2$ is conserved within our system and satisfies the commutation relation $[\hat{N},\hat{H}_2]=0$ when dissipative and driving fields for the atom and cavity are neglected. By fixing the total excitation number, the relevant Hilbert space of atom-cavity system is restricted to the states $|n,b_{1}\rangle,|n-1,b_{2}\rangle,|n-2,b_{3}\rangle,|n-3,b_{4}\rangle$, where $n$ denotes the number of photon excitations. The corresponding matrix is obtained by solving the Schr$\ddot{o}$dinger equation $\hat{{\cal H}}\Psi=M\Psi$, utilizing the associated basis states $\Psi=[|n,b_{1}\rangle,|n-1,b_{2}\rangle,|n-2,b_{3}\rangle,|n-3,b_{4}\rangle]^T$. Consequently, the matrix $M$ is explicitly represented as
\begin{eqnarray} \label{eq:m}
M=\left(\begin{array}{cccc}
n\Delta_{c} & \sqrt{n}g & 0 & 0\\
\sqrt{n}g & \left(n-1\right)\Delta_{c}+\delta & \sqrt{n-1}g & 0\\
0 & \sqrt{n-1}g & \left(n-2\right)\Delta_{c}+2\delta & \sqrt{n-2}g\\
0 & 0 & \sqrt{n-2}g & \left(n-3\right)\Delta_{c}+3\delta
\end{array}\right).
\end{eqnarray}

After the diagonalization matrix~(\ref{eq:m}), the energy spectrum of $n$th $(n\geqslant3)$ dressed state in the spin-$3/2$ JCM is reclassified into four distinct branches. Specifically, the first dressed state splits into two branches ($|1,+1\rangle$ and $|1,-1\rangle$),  while the second dressed state divides into three branches  ($|2,+1\rangle$,$|2,0\rangle$, and $|2,-1\rangle$), as indicated in Fig.~\ref{fig:model}(c).

Explicitly, the energy eigenvalues for $n$-photon excitations $(n\geqslant3)$ for the fixed $\delta/\Delta_{c}=1$ are given by
\begin{eqnarray} \label{splitting}
E_{n,\pm 1}= & n(\Delta_{c}-\Delta_{n,\pm 1}),\nonumber\\
E_{n,\pm 2}= & n(\Delta_{c}-\Delta_{n,\pm 2}),
\end{eqnarray}
 corresponding the $n$-photon resonance ($\Delta_{n,\pm m}$) of the involved dressed states $|n,\pm m \rangle(m=1,2)$ satisfying 
\begin{eqnarray}
\Delta_{n,\pm 1}= &\mp\sqrt{\frac{3n-3-\sqrt{5n^{2}-10n+9}}{2n^2}}g,\nonumber \\ 
\Delta_{n,\pm 2}= &\mp\sqrt{\frac{3n-3+\sqrt{5n^{2}-10n+9}}{2n^2}}g. \nonumber
\end{eqnarray}
here $|n, \pm2\rangle$ index the highest and lowest branches of the splitting states, while $|n, \pm1\rangle$ index the higher and lower two branches.  The energies $E_{n,\pm m}(m=1,2)$ correspond to the states $|n,\pm m \rangle$.  In Fig.~\ref{fig:model}(c), we present the typical anharmonicity ladder of the energy spectrum for spin-$3/2$ JCM. Notably, the asymmetric energy splittings of the $n$th dressed states ($n>2$) for the lower $|n, m=-1\rangle$  and lowest $|n, m=-2\rangle$ branches are labeled as $n\Delta_{n,-1}$ and $n\Delta_{n,-2}$, respectively.

 Figure~\ref{fig:model}(e) characterizes the $n$-photon resonance $\Delta_{n,-m}(m=1,2)$ as a function of $\delta$. As can be seen, the $n$-photon resonance gradually decreases as $\delta$ increases until it converges at $\delta/\Delta_c=1$ with the fixed number of cavity excitations. Additionally, when fixing $\delta/\Delta_c$, the value of single-photon resonance is bigger than the two-photon resonance, and the two-photon resonance outweighs the three-photon resonance  ($\Delta_{1,-1}>\Delta_{2,-1}>\Delta_{3,-2}$), which agrees well with the values in Fig.~\ref{fig:model}(d). We should note that a smaller detuning ratio $\delta/\Delta_c$ results in a considerably larger $n$-photon dressed state splitting. Explicitly, the $n$th dressed state splitting exhibits a symmetrical structure manifesting $n$-photon resonance, satisfying $\Delta_{n,+2}=-\Delta_{n,-2}$.

\section{Simulation method}
\label{simulation method}
To investigate the relevant quantum properties of the spin-3/2 JCM system, including the drive and dissipation of the cavity field and the atom pump field~\cite{PhysRevLett.118.133604}, we calculate the steady state solution by numerically solving the quantum master equation using the Quantum Optics Toolbox~\cite{SzeMTan}. Specifically, the time evolution of the density matrix $\rho$ is governed by the master equation, $d \rho/ d t=\mathcal{L} \rho$. The Liouvillian superoperator $\mathcal{L}$ associated with the Lindblad-type master equation is defined as
\begin{eqnarray}\label{eq:master}
\mathcal{L} \rho= & -i[\hat{H}_1,\rho]+\frac{\kappa}{2}\mathcal{D}[\hat{a}]\rho+\frac{\gamma}{2}\mathcal{D}[\hat{F}_{-}]\rho,
\end{eqnarray}
where $\rho$ is density matrix of the atom-cavity system, $\gamma$ is the spontaneous emission rate of the atom, and $\mathcal{D}[\hat{o}]\rho=2\hat{o}\rho\hat{o}^{\dagger}-\hat{o}^{\dagger}\hat{o}\rho-\rho\hat{o}^{\dagger}\hat{o}$ represents the Lindblad type of dissipation. It's worth noting that, for the ground state of atom, the effect of weak pure dephasing can be disregarded.

For the $k$-th order correlation function related to a time interval $\tau$ for $n$-photons, a significant physical quantity in quantum statistics, is described by~\cite{Munoz2014, Deng:21}:
\begin{eqnarray}\label{eq:gn}
g_{n}^{(k)}(\tau_{1},...\tau_{n})= & \frac{\left\langle \prod_{i=1}^{k}[\hat{a}^{\dagger}(\tau_{i})]^{n}\prod_{i=1}^{k}[\hat{a}(\tau_{i})]^{n}\right\rangle }{\prod_{i=1}^{k}\left\langle [\hat{a}^{\dagger}(\tau_{i})]^{n}[\hat{a}(\tau_{i})]^{n}\right\rangle }
\end{eqnarray}
In the realm of individual photons, the general $k$-order correlation function is represented as $g_{1}^{(k)}(\tau)$. Conversely, the quantum characteristic of emitted multiphoton bundles emission are captured by the correlation function for $n$-photons, denoted as $g_{n}^{(k)}(\tau_{1},\ldots,\tau_{n})$~\cite{PhysRev.130.2529}. Additional, the derivation of the multiphoton correlation function can be involved the application of the quantum regression theorem~\cite{Master}.

The quantum properties of $n$-PB are fundamentally characterized by the correlation functions, which are derived from the steady-state density matrix obtained by numerically solving $\mathcal{L}\rho_{s}=0$. Subsequently, the corresponding number of steady-state photons, denoted as $n_{s}=\rm{Tr} (\hat{a}^{\dag}\hat{a}\rho_s)$, can be calculated. The correlation function for $n$-PB~\cite{PhysRevLett.118.133604, PhysRevLett.121.153601} is expected to satisfy the following criteria:
\begin{eqnarray}\label{eq:npb}
g_{1}^{(n)}(0)>1~ {\rm and} ~g_{1}^{(n+1)}(0)<1.
\end{eqnarray}
These conditions reveal the simultaneous presence of $n$th-order super-Poissonian photon statistics and $(n+1)$th-order sub-Poissonian photon statistics. This implies that the excitation of the first $n$ photons will block the transmission of the subsequent $(n+1)$th photon, leading to an orderly output of $n$-photon stream with sub-Poissonian statistics.

For $n$-photon bundles state ($n>1$), two additional conditions, $g_{1}^{(2)}(0)>g_{1}^{(2)}(\tau)$ and $g_{n}^{(2)}(0)<g_{n}^{(2)}(\tau)$, are required to guarantee photon bunching for the isolated photons and photon anti-bunching for the detached $n$-photon bundles~\cite{Deng:21,Munoz2014,PhysRevLett.117.203602}.  Specifically, single-PB is characterized by criteria $g_{1}^{(2)}(0)<1$ and $g_{1}^{(2)}(0)<g_{1}^{(2)}(\tau)$, to guarantee the sub-Poissonian photon statistics and photon antibunching. A low value of $g_{1}^{(2)}(0)$ indicates a higher purity of single-photon emission. It is note that $g_{1}^{(2)}(0)=0$ corresponds to an ideal single photon blockade with complete suppression of multiphoton excitations. Conversely, the criterion $g^{(n)}_1(0) > 1$ (for $n = 2, 3, 4, 5$) is utilized to identify the PIT effect~\cite{ PhysRevLett.121.153601}, demonstrating super-Poissonian photon statistics~\cite{Faraon2008}.

\section{Numerical results}
\label{numerical results}
In this section, we present the results of numerical simulations, illustrating the feasibility of achieving $n$-photon bundles emission in a spin-$3/2$ JCM  featuring an effective $F=3/2$ alkaline-earth metal atom~\cite{mancini2015observation,PhysRevLett.129.083001,sciadv}),  confined in a high-precision cavity in the current experimental setup. Conventionally, we designate the decay rate of the optical cavity as $\kappa=2\pi\times160$ kHz, serving as the representative energy unit of the system~\cite{sciadv}. Consequently, the long-lived decay rate of the atomic ground state can be safely neglected due to experimentally feasible assumption. The coupling strength of the single atomic cavity is considered to be $g/\kappa=20$. Without loss of generality, we adopt $g_{1}^{{2}}(0) < 0.01$  and $g_{n}^{(2)}(0) < 0.01$ as benchmarks for high-quality single-PB and $n$-photon bundle emission. In addition to strong sub-Poissonian statistics with  $g_{n}^{(2)}(0) \ll 1$, the photon antibunching is further verified by $g_{n}^{(2)}(0)< g_{n}^{(2)}(\tau)$  using the quantum regression theorem. The tunable parameters  in our system encompass the light-cavity detuning $\Delta_{c}$, cavity driven strength $\eta$, Rabi frequency of the classical pump field $\Omega$, and linear Zeeman shift $\delta$.

\subsection{Strong single-PB}
\begin{figure}[ht]
\centering
\includegraphics[width=0.7\columnwidth]{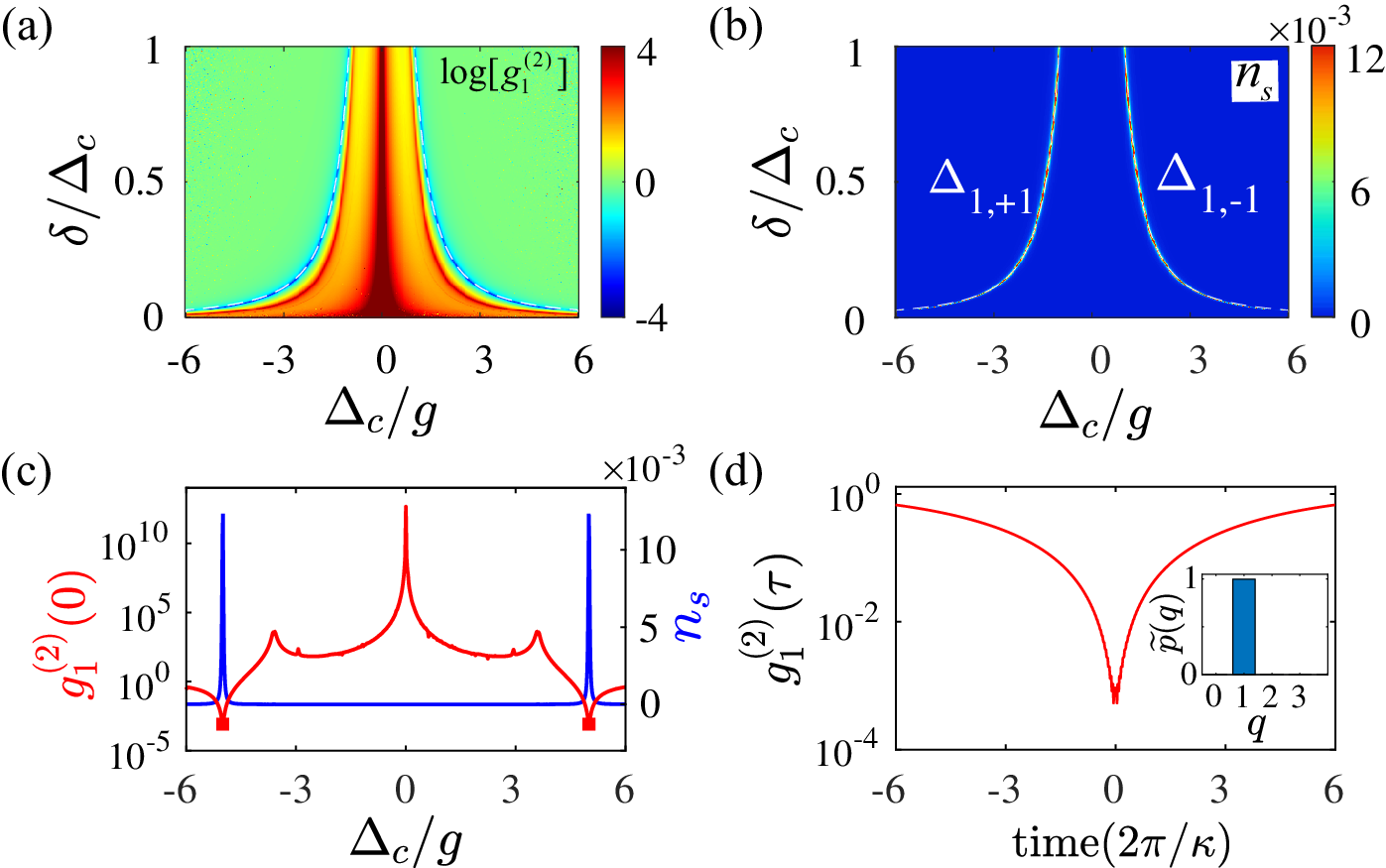}
\caption{(color online) Statistical characteristics of photons in a spin-3/2 JCM with cavity driving. Distributions of (a) equal time second-order correlation function $g_{1}^{(2)}(0)$ and (b) the corresponding $n_{s}$ on the $\Delta_c$-$\delta$ parameter plane.
(c) $g_{1}^{(2)}(0)$ (red line) and $n_{s}$ (blue line) as a function of $\Delta_c$ for $\delta/\Delta_c=0.04$.
(d) Time interval $\tau$ dependence of $g_{1}^{(2)}(\tau)$ for $\delta/\Delta_c=0.04$ and $\Delta_c/g=\pm5$. The inset in (d) depicts the steady-state photon-number distribution $\tilde{p}(q)$ at the single-photon resonances. The other parameters are $\eta/\kappa=0.1$ and $\Omega/\kappa=0$.} \label{fig:1pb}
\end{figure}%

To attain strong single-PB, we initially consider the weak cavity field regime with $\eta/\kappa=0.1$ in the spin-$3/2$ JCM system, while maintaining the pumping field at $\Omega/\kappa=0$. We have checked that the energy spectrum remains almost unchanged due to the sufficiently weak cavity-driven amplitudes ($\eta/g\ll1$ and $\Omega/g\ll1$). Indeed,
$n$-photon resonance is slightly shift even for $\eta/g \sim1$ and $\Omega/g \sim 1$~\cite{deng2021motional}. Figures~\ref{fig:1pb}(a) and \ref{fig:1pb}(b) illustrate the logarithmic plot of the equal-time second-order correlation function $g_{1}^{(2)}(0)$ and the corresponding steady-state cavity photon number $n_{s}$ as functions of the light cavity detuning $\Delta_{c}$ and the linear Zeeman shift $\delta$.
It is evident that both $g_{1}^{(2)}(0)$ and $n_{s}$ exhibit a red-blue symmetric profile with respect to the light-cavity detuning $\Delta_{1,-1}=-\Delta_{1,+1}$ (dashed white lines). Notably, strong single-PB with $g_{1}^{(2)}(0)<0.01$ is realized at the vacuum Rabi splitting $\Delta_{1,\pm 1}$, accompanied by significant steady-state intracavity photon emission.

For a more clearer illustration of the strong single-PB, Figure~\ref{fig:1pb}(c) displays $g_{1}^{(2)}(0)$ and $n_{s}$ versus $\Delta_{c}/g$ with fixing $\delta/\Delta_{c}=0.04$. The photon quantum statistics demonstrate a strong PB with $g_{1}^{(2)}(0)\approx7.3\times10^{-4}$ at the red and blue sidebands of the single-photon resonance $\Delta_{1,\pm 1}/g=\mp5$, corresponding to a larger cavity photon number with $n_{s}\approx0.012$. Particularly, the single-photon resonance points $\Delta_{1,\pm 1}/g=\mp5$ agree well with the analytically derived energy spectrum as displayed in Fig.~\ref{fig:flen}(a) (see \ref{B}). Figure~\ref{fig:1pb}(d) shows the time evolution of the second-order correlation function $g_{1}^{(2)}(\tau)$ at $\delta/\Delta_{c}=0.04$ and the single-photon resonance with $\Delta_{1,-1}/g=5$. Evidently, photon antibunching characterized by $g_{1}^{(2)}(0)<g_{1}^{(2)}(\tau)$ is observable. It's crucial to emphasize that the lifetime of single photon antibunching is dominated by the cavity decay rate, which is proportional to $1/\kappa=6.25$ ms.

To provide further verification of the single-PB characteristic, photon-number distribution is depicted by the ratio of $q$-photon state to the total number of excited photons, i.e., $\tilde{p}(q) = qp(q)/n_s$, where $p(q)={\rm Tr}(|q\rangle\langle q|\rho_s)$ refers to the steady-state photon number distribution  (inset of Fig.~\ref{fig:1pb}(d)). This representation indicates that the photon emission for predominantly comprises single photons,  reaching nearly $100\%$ at $\Delta_{c}/g=\Delta_{1,\pm 1}$. The observed strong single-PB can be attributed to the anharmonicity of the energy spectrum induced by tuning the linear Zeeman shift $\delta$, as illustrated in the ladder diagram of Fig.~\ref{fig:model}(c) for the spin-$3/2$ JCM system. For instance, when the laser field is tuned to resonance with the first excited state ($|1,-\rangle$) at $\delta/\Delta_c$=1,
the system's second energy eigenstate ($|2,-\rangle$) becomes off-resonant, exhibiting an energy gap of $(2-\sqrt{3}/2)g$. This configuration significantly suppresses two-photon excitations, leading to the pronounced manifestation of single-PB effect.

\begin{figure}[ht]
 \centering
\includegraphics[width=0.7\columnwidth]{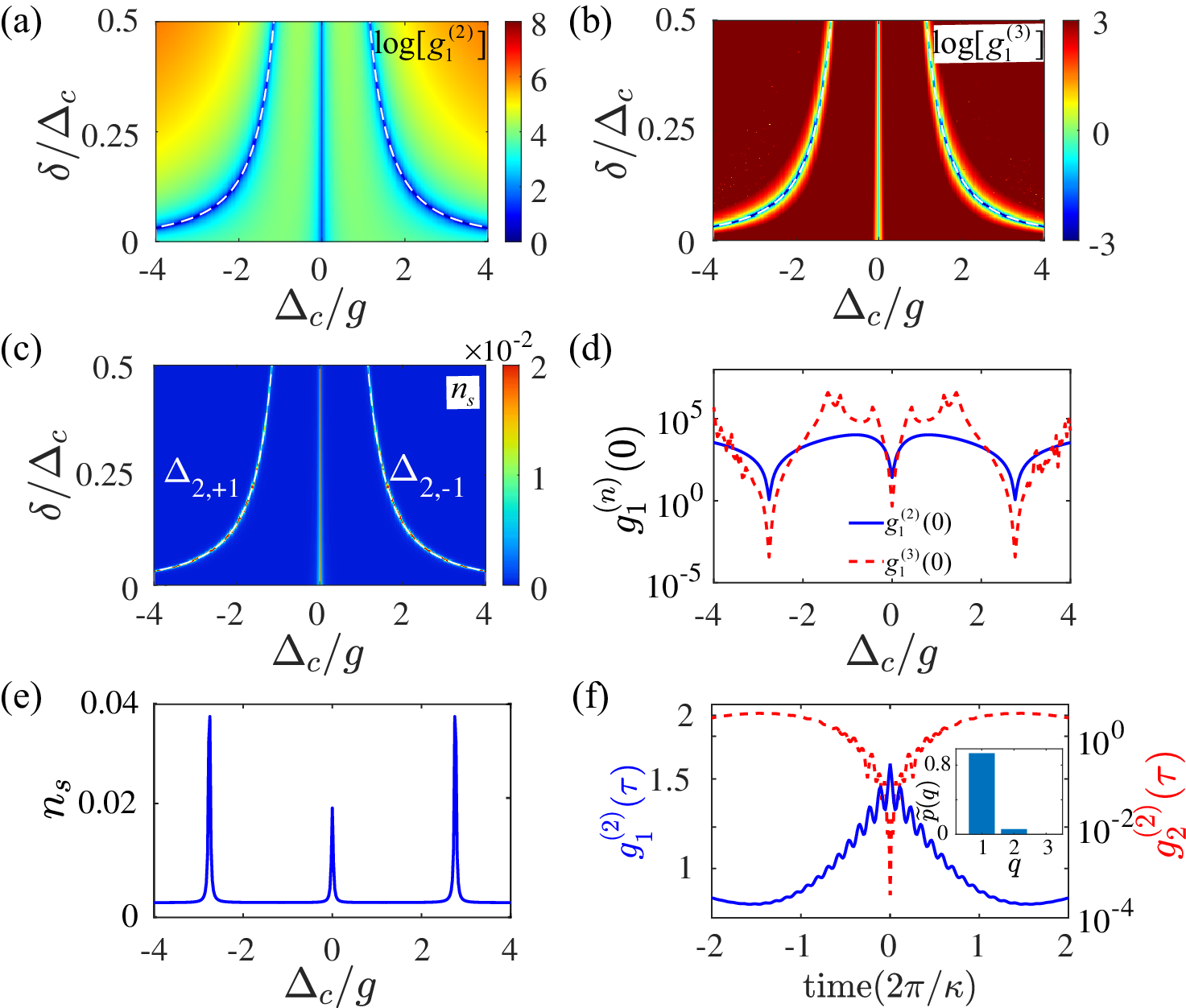}
\caption{(Color online) The statistical properties of photons in a spin-3/2 JCM with atom pump field. Contour plots of (a) log[$g_{1}^{(2)}(0)$], (b) log[$g_{1}^{(3)}(0)$] and (c) $n_{s}$ as functions of $\Delta_c$ and $\delta$. The $\Delta_c$ dependence of (d) $g_{1}^{(2)}(0)$ (blue solid line) and $g_{1}^{(3)}(0)$ (red dashed line), and (e) $n_{s}$ with $\delta/\Delta_c=0.07$. (f) Time interval $\tau$ dependence of $g_{1}^{(2)}(\tau)$ (blue solid line) and $g_{2}^{(2)}(\tau)$  (red dashed line) for $\delta/\Delta_c=0.07$ and $\Delta_c/g=\pm2.7$. The white dashed lines in (a)-(c) show the analytical two-photon resonance $\Delta_{2, \pm 1}$ of the second dressed state in the energy spectrum. Additionally, the inset in (f) plots $\tilde{p}(q)$ at two-photon resonances $\Delta_c/g=\pm2.7$. The other parameters are $\Omega/\kappa=0.05$ and $\eta/\kappa=0$.} \label{fig:2pb}
\end{figure}%
\subsection{High-quality two-photon bundles emission}
In this section, we focus on the quantum characteristics of two-photon bundles emission under the influence of a weak pumping field, with $\Omega/\kappa$ held constant at 0.05 ($\eta/\kappa=0$ ). Figures \ref{fig:2pb}(a)-(c) display the correlation functions $g_{1}^{(2)}(0)$ and $g_{1}^{(3)}(0)$ on a logarithmic scale, as well as the steady-state cavity photon number $n_{s}$, as functions of the detuning $\Delta_{c}$ and the linear Zeeman shift $\delta$. Notably, at the two-photon resonance points $\Delta_{2, \pm 1}$, as indicated by dashed lines, we observe a super-Poissonian distribution ($g_{1}^{(2)}(0)>1$) and a sub-Poissonian distribution ($g_{1}^{(3)}(0)<1$) for the emitted photons, accompanied by significant photon emission. This observation signifies the formation of $2$-PB. Interestingly, the central  branch at $\Delta_{c}=0$ also exhibits substantial photon emission, characterized by $g_{1}^{(2)}(0)>1$ and $g_{1}^{(3)}(0)<1$, confirming the generation of the $2$-PB effect.

Figures \ref{fig:2pb}(d) and \ref{fig:2pb}(e) depict $g_{1}^{(n)}(0)$ (for $n = 2, 3$) and $n_{s}$ as a function of $\Delta_{c}$, with $\delta/\Delta_{c}=0.07$. These figures reveal the implementation of strong $2$-PB, with $g_{1}^{(2)}(0)\approx1.45$ and $g_{1}^{(3)}(0)\approx6.6\times10^{-4}$ at the two-photon resonance points $\Delta_{2,\pm 1}/g=\mp2.7$. Simultaneously, the maximum photon emission $n_{s}$ is observed at the two-photon resonances, serving as another crucial indicator of a high-quality two-photon source. Remarkably, the photon emission of the central branch at $\Delta_{c}=0$ also aligns with the $2$-PB characteristics, exhibiting super-Poissonian statistics ($g_{1}^{(2)}(0)>1$) and sub-Poissonian statistics ($g_{1}^{(3)}(0)<1$) for photon distribution. As shown in the inset of Fig.~\ref{fig:2pb}(f), the two-photon emission nature is evidenced by the photon number distribution $\tilde{p}(q)$, which becomes negligible when $q > 2$. The underlaying mechanism for generating the strong $2$-PB effect mirrors that of single photons, both attributed to the the anharmonicity of the energy spectrum by tuning the linear Zeeman shift in the spin-$3/2$ JCM system. The off-resonant energy gap between the second and fourth energy eigenstates at $\delta/\Delta_c$=1 are $2(\sqrt{3}-\sqrt{2})g$, which suppresses four-photon excitations, leading to the strong $2$-PB effect.
\begin{figure}[ht]
 \centering
\includegraphics[width=0.7\columnwidth]{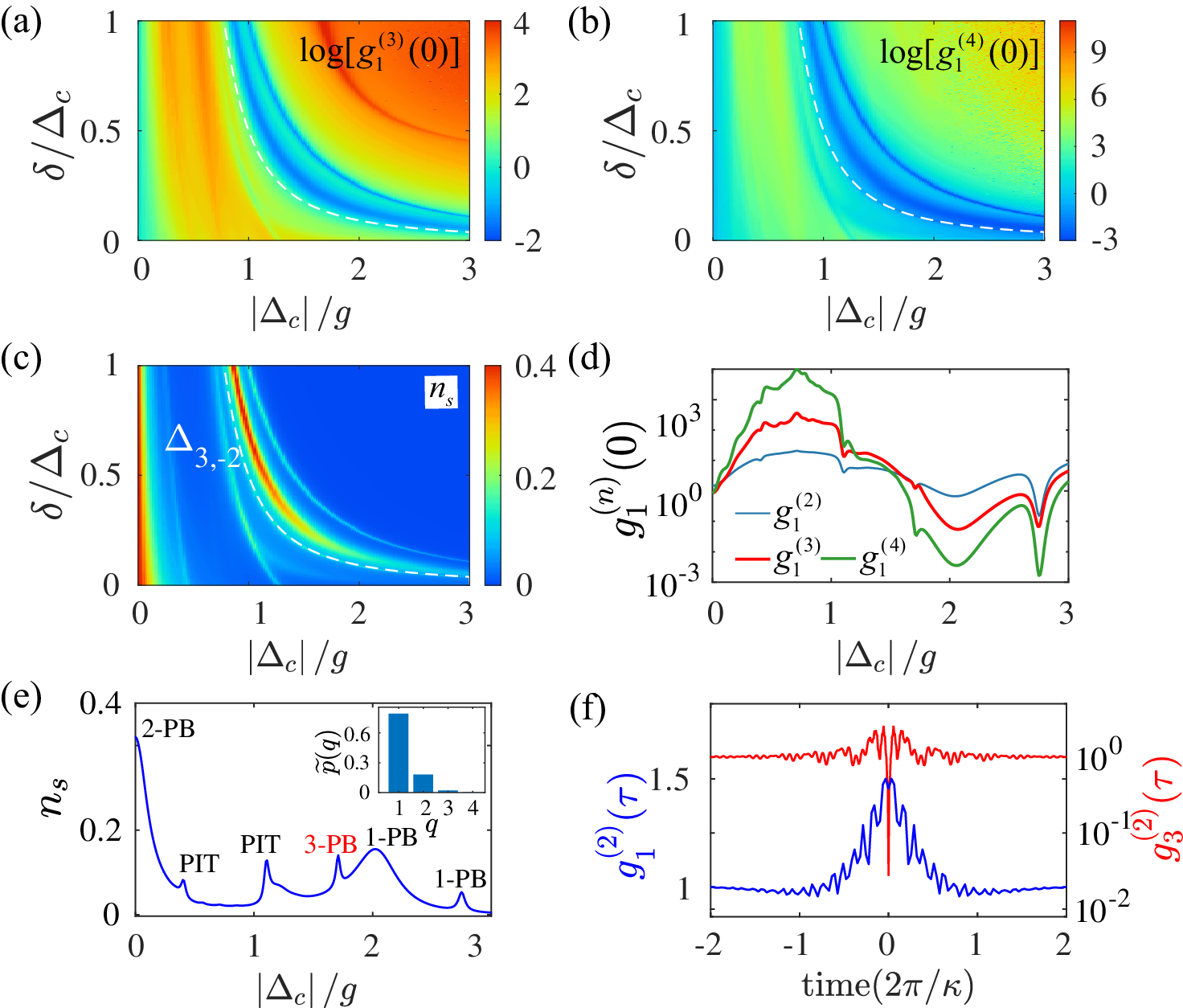}
\caption{(Color online) The statistical characteristics of photons in a spin-3/2 JCM with cavity and atom pump field. Distributions of (a) $g_{1}^{(3)}(0)$, (b) $g_{1}^{(4)}(0)$ and (c) $n_{s}$ on the $\left| \Delta_{c} \right|$-$\delta$ parameter plane.
(d) $g_{1}^{(2)}(0)$ (blue line), $g_{1}^{(3)}(0)$ (red line), $g_{1}^{(4)}(0)$ (green line), and (e) the corresponding $n_{s}$ as a function of  $ |\Delta_{c}|$ for $\delta/\Delta_c=0.13$.
(f) Time interval $\tau$ dependence of $g_{1}^{(2)}(\tau)$ (blue line) and $g_{3}^{(2)}(\tau)$ (red line) for $\delta/\Delta_c=0.13$ and $\Delta_c/g=\pm1.7$. The white dashed lines in (a), (b) and (c) show the analytical three-photon resonance $\Delta_{3, \pm2}$ of the third dressed state of the energy spectrum. The inset in (e) plots $\tilde{p}(q)$ at the three-photon resonances $\Delta_c/g=\pm1.7$. The other parameters are $\eta/\kappa=0.3$ and $\Omega/\kappa=1.3$.} \label{fig:3pb}
\end{figure}%

To further elucidate the nature of two-photon bundles emission, Figure \ref{fig:2pb}(f) shows $g_{1}^{(2)}(\tau)$ and $g_{2}^{(2)}(\tau)$ plotted against the time interval $\tau$ at the two-photon resonance $\Delta_{2, \pm1}/g=\mp2.7$. It is evident that $g_{1}^{(2)}(0)>g_{1}^{(2)}(\tau)$ and $g_{2}^{(2)}(0)<g_{2}^{(2)}(\tau)$, confirming the occurrence of single-photon bunching and two-photon bundles anti-bunching. This result signifies the successful emission of two-photon bundles. Additionally, the decay timescales of single-photon bunching and separated two-photon anti-bunching are comparable, both being proportional to $1/\kappa$.

\subsection{Three-photon bundles emission}
To explore the statistical properties of three-photon bundles emission, we consider both a large cavity drive with $\eta/\kappa=0.3$ and a Rabi frequency of the pump field with $\Omega=1.3$. Notably, Figs.~\ref{fig:3pb}(a)-\ref{fig:3pb}(e) present the quantum statistical properties concerning the red detuning, as both $g_{1}^{(n)}(0)$ and $n_{s}$ exhibit red-blue symmetric structures with the optical cavity detuning.
Figures~\ref{fig:3pb}(a)-\ref{fig:3pb}(c) present the quantum statistics of $\log[g_{1}^{(3)}(0) ]$, $\log[g_{1}^{(4)}(0)]$ and $n_{s}$, respectively, across the $|\Delta_{c}|$-$\delta$ parameter plane. It is observed that at the three-photon resonance $\Delta_{3,-2}/g$ (white dashed line), a three-photon super-Poissonian distribution and a four-photon sub-Poissonian distribution are indicated by $g_{1}^{(3)}(0)>1$ and $g_{1}^{(4)}(0)<1$, alongside the corresponding steady-state photon numbers, demonstrating the creation of $3$-PB.

Figures~\ref{fig:3pb}(d) and~\ref{fig:3pb}(e) demonstrate the values of $g_{1}^{(n)}(0) (n = 2, 3, 4)$ and $n_{s}$ as a function of $\left| \Delta_{c} \right|$ for $\delta/\Delta_{c}=0.13$.  At the three-photon resonance point $\Delta_{3,-2}/g=1.7$, we find $g_{1}^{(3)}(0)=1.35$ and $g_{1}^{(4)}(0)=0.03$, indicating the generation of $3$-PB. Furthermore, significant steady-state photon numbers $n_{s}$ are observed, highlighting the potential for applications in three-photon sources. The three-photon nature is also illustrated by the photon-number distributions $\tilde{p}(q)$, as shown in the inset of Figure~\ref{fig:3pb}(e), exhibiting strong fraction of photon emission from the single-, two-, and three-photon states, while negligible photon emission occurs for $q>3$. Additionally, at the single-photon resonance point $\Delta_{1,-1}/g=2.7$, and the two-photon resonance point $\Delta_{2,-1}/g=2.04$, single-PB effect are observed, with sub-Poissonian photon statistics $g_{1}^{(2)}(0)<1$. Meanwhile, $2$-PB effect is achieved at the two-photon resonance $\Delta_{2,0}/g=0$, realizing $2$-PB with $g_{1}^{(2)}(0)=1.4$ and $g_{1}^{(3)}(0)=0.8$. Interestingly, at the four-photon resonance $\Delta_{4,-2}/g=1.1$, PIT effect is implemented as a manifestation of $g_{1}^{(4)}(0)>g_{1}^{(3)}(0)>g_{1}^{(2)}(0)>1$. These pieces of evidence suggest that a quantum transducer between the single-, $2$-, $3$-PB and PIT effect is achieved. Notably, these resonance points correspond well with specific features in the energy spectrum, as shown in Fig.~\ref{fig:flen}(c) (see \ref{B}).

To further substantiate the characteristics of three-photon bundles emission, Fig.~\ref{fig:3pb}(f) displays $g_{1}^{(2)}(\tau)$ and $g_{3}^{(2)}(\tau)$ at the three-photon resonances $\Delta_{3,\pm2}/g=\mp1.7$. Evidently, single-photon bunching ($g_{1}^{(2)}(0)>g_{1}^{(2)}(\tau)$) and antibunching for isolated three-photon bundles ($g_{3}^{(2)}(0)<g_{3}^{(2)}(\tau)$) are confirmed, indicative of three-photon bundles emission. Importantly, the lifetime of the bunching associated with single photons and the antibunching related to the isolated three-photon bundles share the same timescale.

\subsection{Transducer of three-photon to four-photon bundles emission}
\begin{figure}[ht]
\centering
\includegraphics[width=0.7\columnwidth]{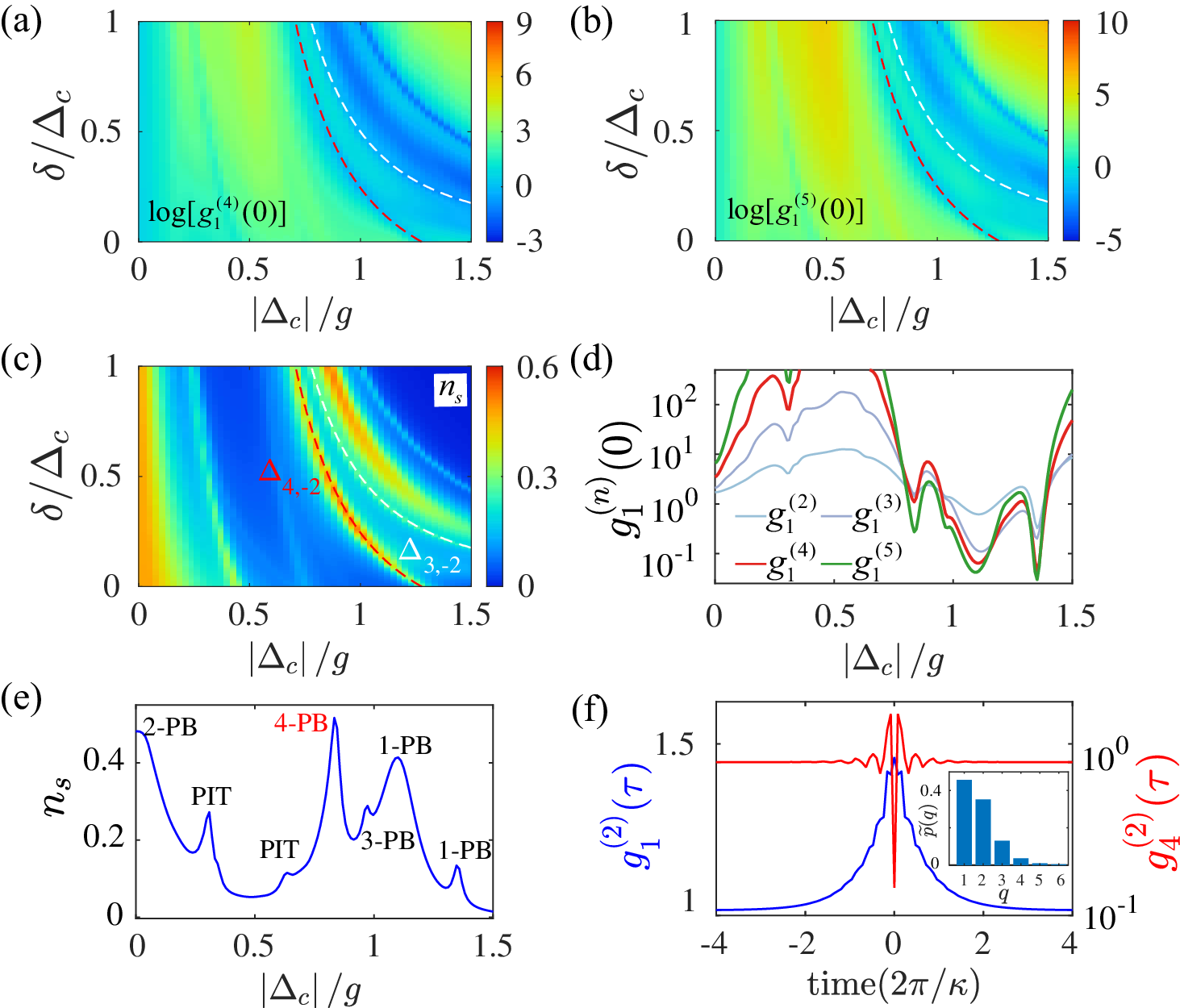}
\caption{(Color online) Investigation of the transduction process from three-photon to four-photon bundles emission in a spin-3/2 JCM. Logarithmic plots depicting (a) $g_{1}^{(4)}(0)$, (b) $g_{1}^{(5)}(0)$ and (c) $n_{s}$ as functions of $\left| \Delta_{c} \right|$ and $\delta$. (d) and (e) showcase the correlation function $g_{1}^{(2)}(0)$, $g_{1}^{(3)}(0)$, $g_{1}^{(4)}(0)$ (red line), and $g_{1}^{(5)}(0)$ (green line), alongside the corresponding photon number $n_{s}$ as a function of $\left| \Delta_{c} \right|$ with $\delta/\Delta_c=0.54$. (f) Time interval $\tau$ dependence of $g_{1}^{(2)}(\tau)$ (blue line) and $g_{4}^{(2)}(\tau)$ (red line) for $\delta/\Delta_c=0.54$ and $\Delta_c/g=\pm0.84$. Insets in (f) plots the steady-state photon-number distribution $\tilde{p}(q)$ at the four photon resonances $\Delta_c/g=\pm0.84$.
The white and red lines in (a), (b), and (c) represent the three-photon ($\Delta_{3,-2}$) and four-photon resonances ($\Delta_{4,-2}$) for the third and fourth dressed states of the energy spectrum, respectively. The other parameters are $\Omega/\kappa=2.3$ and $\eta/\kappa=0.3$.} \label{fig:4pb}
\end{figure}

To probe the multiphoton blockade, we integrate a cavity drive with amplitude $\eta/\kappa=0.3$ and an atomic pump field with Rabi frequency $\Omega/\kappa=2.3$. For enhanced clarity, we emphasize the statistical properties of photons, concerning red detuning, given the red-blue symmetric nature of both $g_{1}^{(n)}(0)$ and $n_s$ relative to the cavity detuning. Figures~\ref{fig:4pb}(a)-(c) depict numerical findings for $g_{1}^{(4)}(0)$, $g_{1}^{(5)}(0)$, and $n_s$ as functions of  $\left| \Delta_{c} \right|$ and $\delta$. Notably, we discern super-Poissonian statistics for four-photons ($g_{1}^{(4)}(0)>1$) and sub-Poissonian statistics for five-photons ($g_{1}^{(5)}(0)<1$) at the four-photon resonance $\Delta_{4,-2}$ (indicated by red dashed line), indicating successful generation of 4-PB. The corresponding region also shows a significant photon number, as illustrated in Fig.~\ref{fig:4pb}(c).

To enhance our understanding of 4-PB generation, we depict $n_s$ and $g_{1}^{(n)}(0) (n = 2, 3, 4, 5)$ as functions of $\left| \Delta_{c} \right|$, with $\delta/\Delta_c=0.54$,  as shown in Figs.~\ref{fig:4pb}(d) and~\ref{fig:4pb}(e). At the four-photon resonance with $\Delta_{4,- 2}/g=0.84$, we confirm the generation of 4-PB, evidenced by $g_{1}^{(4)}(0) = 1.49$ and $g_{1}^{(5)}(0) = 0.3$. The four-photon emission characteristics are further validated by the phonon-number distribution $\tilde{p}(q)$, wherein negligible small values for $q>4$, possessing the similar statistical properties observed in three-photon emission, as shown in the inset of Fig.~\ref{fig:4pb}(f). Simultaneously, 3-PB is achieved at the three-photon resonance with $\Delta_{3,-2}/g=0.98$, displaying three-photon super-Poissonian statistics with $g_{1}^{(3)}(0)=1.5$ and four-photon sub-Poissonian statistics with $g_{1}^{(4)}(0)=0.5$. Intriguingly, PIT emerges at $ \Delta_{c} /g=0$ and $\left| \Delta_{c} \right|/g=0.31$, showcasing the hierarchy $g_{1}^{(5)}(0)>g_{1}^{(4)}(0)>g_{1}^{(3)}(0)>g_{1}^{(2)}(0)>1$. Conversely, single-PB manifests at both the two-photon resonance $\Delta_{2,-1}/g=1.1$ and the single-photon resonance $\Delta_{1,-1}/g=1.3$, exhibiting the sub-Poissonian statistics $g_{1}^{(2)}(0)<1$. These results denote a notable progression in the exploration of non-classical photon emission, demonstrating a transducer among single-PB, PIT, 3-PB and 4-PB through the manipulation of cavity-light detuning.

Figure~\ref{fig:4pb}(f) further explores the interval dependence of  correlation functions $g_{1}^{(2)}(\tau)$ and $g_{4}^{(2)}(\tau)$ at the four-photon resonance $\Delta_{4,\pm 2}/g= \mp 0.84$, confirming single-photon bunching ($g_{1}^{(2)}(0)>g_{1}^{(2)}(\tau)$) and four-photon bundles anti-bunching ($g_{4}^{(2)}(0)<g_{4}^{(2)}(\tau)$) with comparable lifetimes proportional to $1/\kappa$. The underlying physical mechanism for the unique non-classical photon emission relies on the high-spin degree-of-freedom atom-cavity system,  which enhances optical nonlinearity and fosters interaction between the cavity and atomic field. Interestingly, our approach facilitates the realization of multimode bundles spanning between three and four photons, thereby enabling higher-order processes to achieve higher-PB. Notably, the inherent capabilities of the high-spin degree of freedom system compensates the strong atom-cavity coupling required by the generation of strong $n$-PB. Our finding unveils promising avenues for studying the special non-classical quantum states, offering profound implications for advancements in quantum computing and communication~\cite{s41586-022-04987-5, Yang2022}.

\section{Conclusions}
\label{conclusion}
Building upon experimental advancements in atomic cavity QEDs, we theoretically explore the emission of $n$-photon bundles from a single spin-$3/2$ atom coupled to a single-mode optical cavity. Our investigation reveals a significant enhancement in the anharmonicity of the energy spectrum, correlated with the splitting of the $n$th dressed state can be significantly enhanced by adjusting the linear Zeeman shift in the spin-$3/2$ JCM. By appropriately driving the optical cavity and atom under the optimal linear Zeeman shift $\delta$, we achieve the high-quality single-photon emission with $g_{1}^{(2)}(0) \approx 7.3 \times 10^{-4}$, as well as two-photon bundles emission with $g_{1}^{(2)}(0)>1$ and  $g_{1}^{(3)}(0) \approx 6.6 \times 10^{-4}$. A pivotal outcome of our investigation is the successful conversion from three-photon bundles to four-photon bundles emission, realized by simultaneously driving both the cavity and atom at suitable atomic-cavity detunings. These results representing a significant advancement is facilitated by the high spin degree of freedom, which compensates for the strong nonlinear interactions  in our spin-$3/2$ JCM framework. Moreover, our study demonstrates the potential for obtaining high-quality multi-photon sources with correspondingly large steady-state photon numbers. This study provides an exceptional framework for exploring novel quantum states by harnessing spin degrees of freedom in high-spin systems, offering promising prospects for the development of non-classical quantum switches, multimode bundle splitters, and quantum information processing applications.

\section*{Acknowledgments}

This work was supported by the National Natural Science Foundation of China (Grant No.12374365, Grant No. 12274473, and Grant No. 12135018) and the Fundamental Research Funds for the Central Universities (Grant No. 24qnpy120).

\appendix

\section{Hamiltonian for spin-3/2 JCM}
\label{A}
In this section we derive the Hamiltonian~(\ref{eq:H1}) based on the energy level structure and laser configuration presented in Fig.~\ref{fig:model} in the main text.
Under the rotation approximation, the Hamiltonian of a single atom cavity
coupled system is expressed as
\begin{eqnarray}\label{eq:flh1}
\hat{\cal{H}}_1/\hbar&=\omega_{c}\hat{a}^{\dagger}\hat{a}+\sum \limits _{j=1} ^{3} j\omega_{b}\hat{b}_{j+1}^{\dagger}\hat{b}_{j+1}
+\sum \limits _{j=1} ^{3} \left(\omega_{e}+j\omega_{b}'\right)\hat{e}_{j}^{\dagger}\hat{e}_{j} \nonumber \\
		& + \left[g_{1}\hat{a} \sum \limits _{j=1} ^{3}\hat{e}_{j}^{\dagger}\hat{b}_{j}
		 +\Omega_{1} \sum \limits _{j=1} ^{3} \hat{e}_{j}^{\dagger}\hat{b}_{j+1}e^{-i\omega_{L}t}+\rm H.c.\right],
\end{eqnarray}
where $\hat{a}^{\dagger}$ ($\hat{a}$) is the creation (annihilation)
operator for the cavity mode and $\hbar\omega_{b}$ ($\hbar\omega_{b}'$) represents the linear Zeeman shift within the ground (excited) state manifold. The annihilation operators for the three excited states and four ground states are respectively defined as $\hat{e}_j$ with $j=1,2,3$ and $\hat{b}_j$ with $j=1,2,3,4$. To eliminate time dependence, we define
the unitary transformation as 
\begin{eqnarray}
{\cal U}=\exp\left[{-i}\omega_{L}t\left(\hat{a}^{\dagger}\hat{a}+\sum \limits _{j=1} ^{3}\hat{e}_{j}^{\dagger}\hat{e}_{j}\right)\right],
\label{eq:u}
\end{eqnarray}
then the Hamiltonian~(\ref{eq:flh1}) can be represented as
\begin{eqnarray}
\hat{\cal H}_2/\hbar&={\cal U}^{\dagger}\hat{\cal H}_1{\cal U}-i{\cal U}^{\dagger}\frac{\partial}{\partial t}{\cal U}\nonumber \\
&=	\Delta_{c}^{'}\hat{a}^{\dagger}\hat{a}+\sum \limits _{j=1} ^{3} j\omega_{b}\hat{b}_{j+1}^{\dagger}\hat{b}_{j+1}+\Delta \sum \limits _{j=1} ^{3}\hat{e}_{j}^{\dagger}\hat{e}_{j}   \nonumber \\
		& + \left[g_{1}\hat{a} \sum \limits _{j=1} ^{3} \hat{e}_{j}^{\dagger}\hat{b}_{j}
		 +\Omega_{1} \sum \limits _{j=1} ^{3} \hat{e}_{j}^{\dagger}\hat{b}_{j+1}+\rm H.c.\right],
\label{eq:flh2}
\end{eqnarray}
where $\Delta_{c}^{'}=\omega_{c}-\omega_{L}$ is the light-cavity
detuning, $\Delta=\omega_{e}+j\omega_{b}'-\omega_{L} (j=1,2,3)$
is the one-photon detuning of the atom. To eliminate the three excited states $|e_j\rangle$ with $j=1,2,3$ the Heisenberg equations of motion for atomic operators can be written as
\begin{eqnarray}
i\frac{d\hat{e}_{j}}{dt}= &  \Delta\hat{e}_{j}+g_{1}\hat{a}\hat{b}_{j}+\Omega_{1}\hat{b}_{j+1} (j=1,2,3),\nonumber \\
i\frac{d\hat{b}_{1}}{dt}= &   g_{1}\hat{a}^{\dagger}\hat{e}_{1},\nonumber \\
i\frac{d\hat{b}_{2}}{dt}= &   \omega_{b}\hat{b}_{2}+g_{1}\hat{a}^{\dagger}\hat{e}_{2}+\Omega_{1}\hat{e}_{1},\nonumber \\
i\frac{d\hat{b}_{3}}{dt}= &   2\omega_{b}\hat{b}_{3}+g_{1}\hat{a}^{\dagger}\hat{e}_{3}+\Omega_{1}\hat{e}_{2},\nonumber \\
i\frac{d\hat{b}_{4}}{dt}= &   3\omega_{b}\hat{b}_{4}+\Omega_{1}\hat{e}_{3}.
\label{eq:eb}
\end{eqnarray}

Subsequently, in the dispersive regime $\Delta\gg\{g_{1},\Omega_{1}\}$, the atomic
excited states $|e_j\rangle$ with $j=1,2,3$
can be adiabatically eliminated, i.e., $i{d\hat{e}_j}/{dt}=0$, which yields
\begin{eqnarray}
\hat{e}_{j}= &  -\frac{g_{1}\hat{a}\hat{b}_{j}+\Omega_{1}\hat{b}_{j+1}}{\Delta},(j=1,2,3).
\label{eq:e}
\end{eqnarray}
Substituting Eq.~(\ref{eq:e}) into Eq.~(\ref{eq:eb}), the Heisenberg equations with respect to the ground state operators are denoted by
\begin{eqnarray}
i\frac{d\hat{b}_{1}}{dt}= &   -\frac{g_{1}^2\hat{a}^{\dagger}\hat{a}\hat{b}_1+g_{1}\Omega_{1}\hat{a}^{\dagger}\hat{b}_2}{\Delta},\nonumber \\
i\frac{d\hat{b}_{2}}{dt}= &   \omega_{b}\hat{b}_{2}-\frac{(g_{1}\hat{a}^{\dagger}\hat{a}+|\Omega_{1}|^2)\hat{b}_2+g_{1}\Omega_{1}\hat{a}^{\dagger}\hat{b}_3
+g_{1}\Omega_{1}\hat{a}\hat{b}_1}{\Delta},\nonumber \\
i\frac{d\hat{b}_{3}}{dt}= &   2\omega_{b}\hat{b}_{3}-\frac{(g_{1}\hat{a}^{\dagger}\hat{a}+|\Omega_{1}|^2)\hat{b}_3+g_{1}\Omega_{1}\hat{a}^{\dagger}\hat{b}_4
+g_{1}\Omega_{1}\hat{a}\hat{b}_2}{\Delta},\nonumber \\
i\frac{d\hat{b}_{4}}{dt}= &   3\omega_{b}\hat{b}_{4}-\frac{g_{1}\Omega_{1}\hat{a}\hat{b}_3+|\Omega_{1}|^2\hat{b}_4}{\Delta}.
\label{eq:b}
\end{eqnarray}

Accordingly, the Hamiltonian~(\ref{eq:flh2}) reduces to
\begin{small}
\begin{eqnarray}
{\cal H}_3/\hbar&= \Delta_{c}\hat{a}^{\dagger}\hat{a}+\Delta_{a}\sum \limits _{j=1} ^{3} j \hat{b}_{j+1}^{\dagger}\hat{b}_{j+1}
+\left[ g\hat{a}^{\dagger}\sum \limits _{j=1} ^{3}\hat{b}_{j}^{\dagger}\hat{b}_{j+1}+\rm H.c.\right],
\label{eq:flh3}
\end{eqnarray}
\end{small}
where $\Delta_{c}=\Delta_{c}^{'}-g_{1}^2/\Delta$ represents the effective light-cavity detuning, and $\Delta_{a}=\omega_{b}-|\Omega_{1}|^2/\Delta$ indicates the effective single-photon detuning. In addition, $g=-g_{1}\Omega_{1}/\Delta$ is the effective single atom-cavity coupling strength.

Furthermore, taking into account the driving terms of the cavity field and the atom field, the time-independent driven Hamiltonian under the unitary transformation ({\ref{eq:u}}) can be written as ${\cal H}_d=\eta(\hat{\emph a}^{\dagger}+\hat{\emph a})+\Omega\sum \limits_{\emph j=1} ^{2}\hat{b}_{j}^{\dagger}\hat{b}_{j+2}+\rm H.c.$. As a result, the total Hamiltonian of the single atom-cavity is given by
\begin{eqnarray}
{\cal H}_4/\hbar&= \Delta_{c}\hat{a}^{\dagger}\hat{a}+\Delta_{a}\sum \limits _{j=1} ^{3} j \hat{b}_{j+1}^{\dagger}\hat{b}_{j+1}+ \eta(\hat{\emph a}^{\dagger}+\hat{\emph a}) \nonumber
\\& +\left[ g\hat{a}^{\dagger}\sum \limits _{j=1} ^{3}\hat{b}_{j}^{\dagger}\hat{b}_{j+1}+\Omega\sum \limits_{\emph j=1} ^{2}\hat{b}_{j}^{\dagger}\hat{b}_{j+2}+\rm H.c.\right],
\label{eq:flh4}
\end{eqnarray}
the Hamiltonian (\ref{eq:flh4}) for the spin-3/2 JCM system is the identical to Eq. (\ref{eq:H1}) in the main text.

\section{Eigenenergy spectrum of atom-cavity system}
\label{B}

To deeper our understanding of the spin-$3/2$ JCM system, we delve into its energy spectrum. Under the scenario of weakly driven atom and cavity fields, where the total number of excitations remains constant. The Hilbert space is constrained to four Fock basis states denoted as $|n, b_{1}\rangle,|n-1,b_{2}\rangle,|n-2,b_{3}\rangle$ and
$|n-3,b_{4}\rangle$, each associated with the excitation number $n$. The matrix $M$ of diagonalizing the Hamiltonian (\ref{eq:H2}) in the main text under the $n\geqslant3$ subspace is expressed as
\begin{eqnarray}
\left(\begin{array}{cccc}
n\Delta_{c} & \sqrt{n}g & 0 & 0\\
\sqrt{n}g & \left(n-1\right)\Delta_{c}+\delta & \sqrt{n-1}g & 0\\
0 & \sqrt{n-1}g & \left(n-2\right)\Delta_{c}+2\delta & \sqrt{n-2}g\\
0 & 0 & \sqrt{n-2}g & \left(n-3\right)\Delta_{c}+3\delta
\end{array}\right).
\end{eqnarray}

It can be anticipated that the respective $n\geqslant3$ dressed states split into four branches in the pseudospin-$3/2$ manifold. In order to simplify the derivation of analytical solutions, we examine the characteristic eigenenergy spectrum of the system by setting $\delta= \Delta _c$. These energy eigenvalues are represented as:
\begin{eqnarray}
E_{n,\pm 1}= & n(\Delta_{c}-\Delta_{n,\pm 1}),\nonumber\\
E_{n,\pm 2}= & n(\Delta_{c}-\Delta_{n,\pm 2}),\nonumber\\
\Delta_{n,\pm 1}= & C_{n,\pm 1}g,\nonumber\\
\Delta_{n,\pm 2}= & C_{n,\pm 2}g,\nonumber\\
C_{n,\pm 1}=&\mp\sqrt{\frac{3n-3-\sqrt{5n^{2}-10n+9}}{2n^2}},\nonumber\\
C_{n,\pm 2}=&\mp\sqrt{\frac{3n-3+\sqrt{5n^{2}-10n+9}}{2n^2}}.
\label{eq:en}
\end{eqnarray}
here $|n, \pm2\rangle$ corresponds to the highest and lowest branches of the energy splitting of the dressed states, and $|n, \pm1\rangle$ denote higher and lower branches. Especially, the energy splitting between the dressed states $|3, -2\rangle$ and $|3, +2\rangle$ is $2\sqrt{3+\sqrt{6}}g$, as depicted in Fig.~\ref{fig:model}(c). Specifically, the representation of the dressed states is given by
\begin{eqnarray}
\left|n, m\right\rangle  & =\sin\theta_{m} \cos\varphi_{m} \sin\lambda_{m} \left|n,b_{1}\right\rangle\nonumber \\
 & +\sin\theta_{m} \sin\varphi_{m}\sin\lambda_{m}\left|n-1,b_{2}\right\rangle\nonumber \\
 & +\cos\theta_{m} \sin\lambda_{m}\left|n-2,b_{3}\right\rangle\nonumber \\
 & +\cos\lambda_{m}\left|n-3,b_{4}\right\rangle (m=\pm1,\pm2).
\end{eqnarray}

\begin{figure}[ht]
\centering
\includegraphics[width=0.7\textwidth]{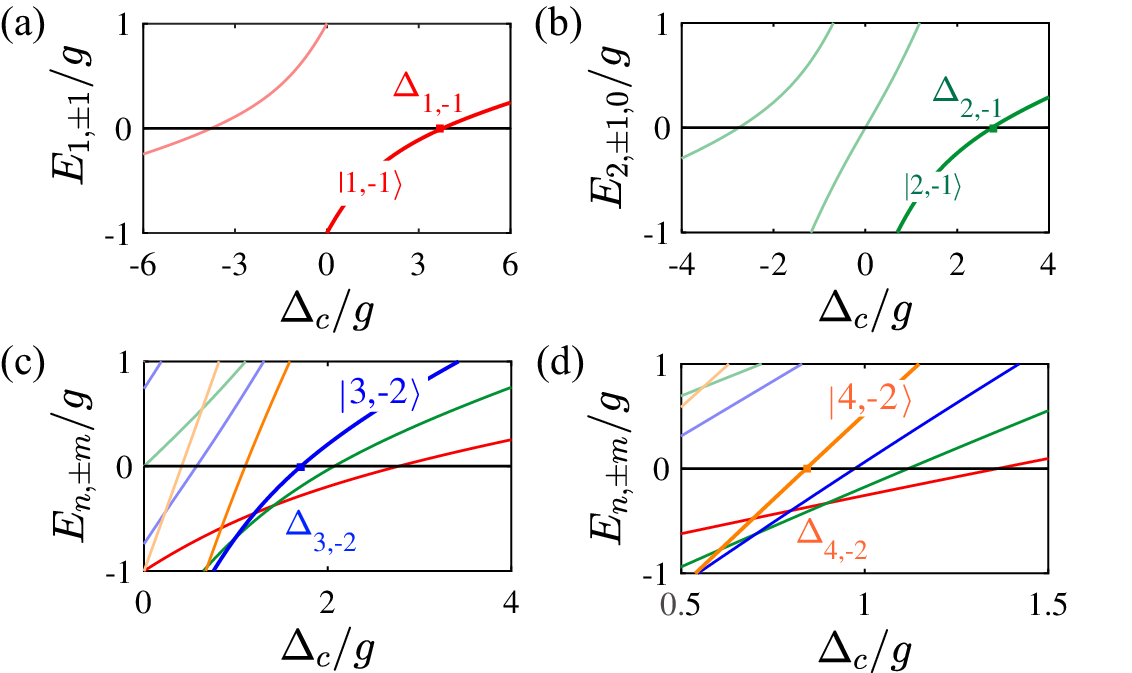}
\caption{(Color online) The typical  energy spectrum for (a) $\delta/\Delta_c=0.04$, (b) $\delta/\Delta_c=0.07$, (c) $\delta/\Delta_c=0.13$, and (d) $\delta/\Delta_c=0.54$.}
\label{fig:flen}
\end{figure}

The representation of the dressed states involves mixed angles $\varphi_{m}$, $\theta_{m}$, and $\lambda_{m}$, which determine the population distribution among the different Fock states. Three mixed angles satisfy the following conditions
\begin{eqnarray}
\tan\varphi_{m}=&\frac{C_{n, m}}{\sqrt{n}}, \nonumber \\
\tan\theta_{m}=&\frac{\sqrt{(n-1)(n+C^{2}_{n, m})}}{{C^{2}_{n, m}}-n},\nonumber \\
\tan\lambda_{m}=&\frac{C_{n, m}\sqrt{(n-1)(n+C^{2}_{n, m})+(C^{2}_{n, m}-n)^{2}}}{{\sqrt{(n-2)}(C^{2}_{n, m}-n})}.
\end{eqnarray}
With $m=\pm1,\pm2$.
Figure~\ref{fig:flen} illustrates the energy spectrum of Hamiltonian (\ref{eq:H2}) for various specific parameters. The resonance points in the energy spectrum correspond respectively to single-PB, two-photon bundles emission, three-photon bundles emission, and four-photon bundles emission mentioned in the main text.

Regarding two-photon excitation, the matrix of Hamiltonian (\ref{eq:H2}) can be reduced to
\begin{eqnarray}
\left(\begin{array}{ccc}
n\Delta_{c} & \sqrt{n}g & 0\\
\sqrt{n}g & \left(n-1\right)\Delta_{c}+\delta & \sqrt{n-1}g\\
0 & \sqrt{n-1}g & \left(n-2\right)\Delta_{c}+2\delta
\end{array}\right).
\label{eq:2ppb}
\end{eqnarray}

Deterministically, the dressed state splits into three branches following diagonalising the Hamiltonian~(\ref{eq:2ppb}). The properties of the eigenenergy spectrum can be investigated more efficiently by fixing $\delta= \Delta _c$. These eigenenergy are presented by
\begin{eqnarray}
E_{2,\pm 1}= & 2\Delta_{c}\mp\sqrt{3}g,\nonumber \\
E_{2,0}= & 2\Delta_{c}.
\end{eqnarray}
As a result, the two photon resonances can be expressed as $\Delta_{2,\pm 1}=\mp \sqrt{3}g/2$ and $\Delta_{2,0}=0$. The energy splitting between the dressed states $|2,-1\rangle$ and $|2,0\rangle$ is given by $2\Delta_{2,-1}=\sqrt{3}g$. Conversely, the energy splitting between the dressed states $|2,-1\rangle$ and $|2,+1\rangle$ amounts to $2\sqrt{3}g$, as shown in Fig~\ref{fig:model}(c).

The corresponding eigenstates read
\begin{eqnarray}
\left|2,+1\right\rangle  & =-\sin\theta \cos\varphi\left|2,b_{1}\right\rangle +\sin\varphi\left|1,b_{2}\right\rangle\nonumber \\
 & -\cos\theta \cos\varphi\left|0,b_{3}\right\rangle \nonumber  \\
\left|2,0\right\rangle &=\cos\theta\left|2,b_{1}\right\rangle -\sin\theta\left|0,b_{3}\right\rangle\nonumber  \\
\left|2,-1\right\rangle  & =\sin\theta \sin\varphi\left|2,b_{1}\right\rangle +\cos\varphi\left|1,b_{2}\right\rangle\nonumber \\
 & +\cos\theta \sin\varphi\left|0,b_{3}\right\rangle
\end{eqnarray}
where two mixed angles need to satisfy
\begin{eqnarray}
\tan\theta=&\sqrt{2},\nonumber \\
\tan\varphi=&1.
\end{eqnarray}

Concerning the process of single-photon excitation, the matrix of Hamiltonian (\ref{eq:H2}) is reduced to
\begin{eqnarray}
\left(\begin{array}{cc}
n\Delta_{c} & \sqrt{n}g\\
\sqrt{n}g & (n-1)\Delta_{c}+\delta
\end{array}\right).
\label{eq:1p}
\end{eqnarray}

Upon diagonalizing the Hamiltonian~(\ref{eq:1p}), the dressed states are found to bifurcate into two branches, with corresponding eigenenergies given by $E_{1,\pm 1}=\left(\Delta_{c}+\delta\right)/2\pm\sqrt{g^{2}+\left(\delta-\Delta_{c}\right)^{2}/4}$. By fixing $\delta=\Delta_c$ in Fig.~\ref{fig:model} (c), the eigenenergies simplify to $E_{1,\pm 1}=\Delta_c\mp g$. Consequently, the single photon resonance can be characterized by $\Delta_{1,\pm 1}=\mp g$. The energy splitting between the dressed states $|1,-1\rangle$ and $|1,+1\rangle$ is $2g$.

Specifically, the eigenstates can be written as
\begin{eqnarray}
\left|1,+1\right\rangle =\sin\theta\left|1,b_{1}\right\rangle -\cos\theta\left|0,b_{2}\right\rangle, \\ \nonumber
\left|1,-1\right\rangle =\cos\theta\left|1,b_{1}\right\rangle +\sin\theta\left|0,b_{2}\right\rangle,
\end{eqnarray}
where the mixed angle meets the following conditions
\begin{eqnarray}
\tan\theta=\sqrt{\frac{\varepsilon-\left(\delta-\Delta_{c}\right)}{\varepsilon+\left(\delta-\Delta_{c}\right)}},
\end{eqnarray}
with $\varepsilon=\sqrt{(\delta-\Delta_{c})^{2}+4g^{2}}$.

In summary, the anharmonicity energy spectrum can be sketched as shown in Fig.~\ref{fig:model}(c).
\section*{Data availability statement}

All data that support the findings of this study are included within the article (and any supplementary files).

\section*{References}

\end{document}